\documentclass[usenatbib,usegraphicx]{mn2e}
\usepackage{amsmath}
\usepackage[]{hyperref}
\usepackage{graphicx}
\usepackage{color}
\usepackage{units}
\usepackage{ulem}
\usepackage{fixltx2e}
\DeclareGraphicsExtensions{.eps,.pdf,.png,.jpg}

\usepackage[T1]{fontenc}
\usepackage{textcomp}
\usepackage[utf8]{inputenc}

\hypersetup{
    colorlinks=true,
    linkcolor=black,
    citecolor=black,
    filecolor=black,
    urlcolor=black,
}

\begin{document}

\title[PDF PSA]{A PDF PSA, or Never gonna set\_xscale again - guilty feats with logarithms. }

\author[J. C. Forbes]{John C. Forbes$^1$\textsuperscript{\thanks{E-mail: jforbes@flatironinstitute.org}} \\
$^1$Flatiron Institute, Simons Foundation, 162 Fifth Avenue, New York, NY 10010\\
*E-mail: jforbes@flatironinstitute.org
}

\maketitle

\begin{abstract}
In the course of doing astronomy, one often encounters plots of densities, for example probability densities, flux densities, and mass functions. Quite frequently the ordinate of these diagrams is plotted logarithmically to accommodate a large dynamic range. In this situation, I argue that it is critical to adjust the density appropriately, rather than simply setting the x-scale to `log' in your favorite plotting code. I will demonstrate the basic issue with a pedagogical example, then mention a few common plots where this may arise, and finally some possible exceptions to the rule.
\end{abstract}

\section{Introduction}

Astronomers study, among other things, large populations of objects, spectra of distant objects, and models with physically meaningful parameters that may be inferred through Bayesian inference. As such it is often convenient in the course of doing astronomy to plot probability density functions (PDF's) and closely related quantities.

Although there are philosophical disagreements on the definition of probability, it is uncontroversial that probability is a dimensionless number between 0 and 1. Probability densities, on the other hand, are not dimensionless. They have units of the inverse of the random variable over which they are defined. This immediately follows from dimensional analysis of the requirement that the integral of a PDF over the domain of the random variable is 1, i.e. for a PDF $p(x)$
\begin{equation}
\int_{-\infty}^\infty p(x) dx = 1
\end{equation}
Despite this almost trivial fact, astronomers have a terrible habit of labelling the $y-$axis in a plot of PDFs simply ``PDF'' or ``Probability.'' The latter is plainly wrong. A PDF and a probability are fundamentally different things, which will generally have different units. 

Labelling an axis simply ``PDF'' is somewhat better, but is likely to be ambiguous. The PDF of what? The x-coordinate? A common source of ambiguity is the fact that the $x-$coordinate is often plotted logarithmically since the data extend over orders of magnitude in that quantity, e.g. mass, wavelength, or luminosity. If the $y-$ axis is labelled simply ``PDF,'' it is not clear if it is the PDF of the $x$-variable or of $\log_{10} x$.

If the y-axis is in fact the PDF of $x$, not $\log_{10} x$, but the x-axis is logarithmic, this creates an additional problem beyond whether or not the $y-$axis is correctly labelled. In this situation, the plot itself is {\bf highly misleading}. This stems from the fact that the area under the curve being plotted no longer represents the probability in that $x-$range! To put it explicitly, in general
\begin{equation}
\label{eq:integral}
\int_a^b p(x) dx \ne \int_{\log_{10}a}^{\log_{10}b} p(10^u) du,
\end{equation}
where $u = \log_{10} x$, because the change of variables requires substituting $dx = \ln(10) x du$, not just $du$. However, when plotting the PDF of $x$ with respect to a logarithmically-scaled $x-$axis, the area under the curve is in fact the incorrect right-hand side of Equation \eqref{eq:integral}.



\section{Generic Pedagogical Example}
To illustrate the argument, I'm going to look in some detail at a density function which is constant in log-space, but of course behaves quite differently in linear space. I will then plot the PDF of this variable in a variety of ways. I argue that some of them, which are common in astronomy, are misleading if not outright wrong.

\begin{figure*}
\centering
\includegraphics[width=7.5in]{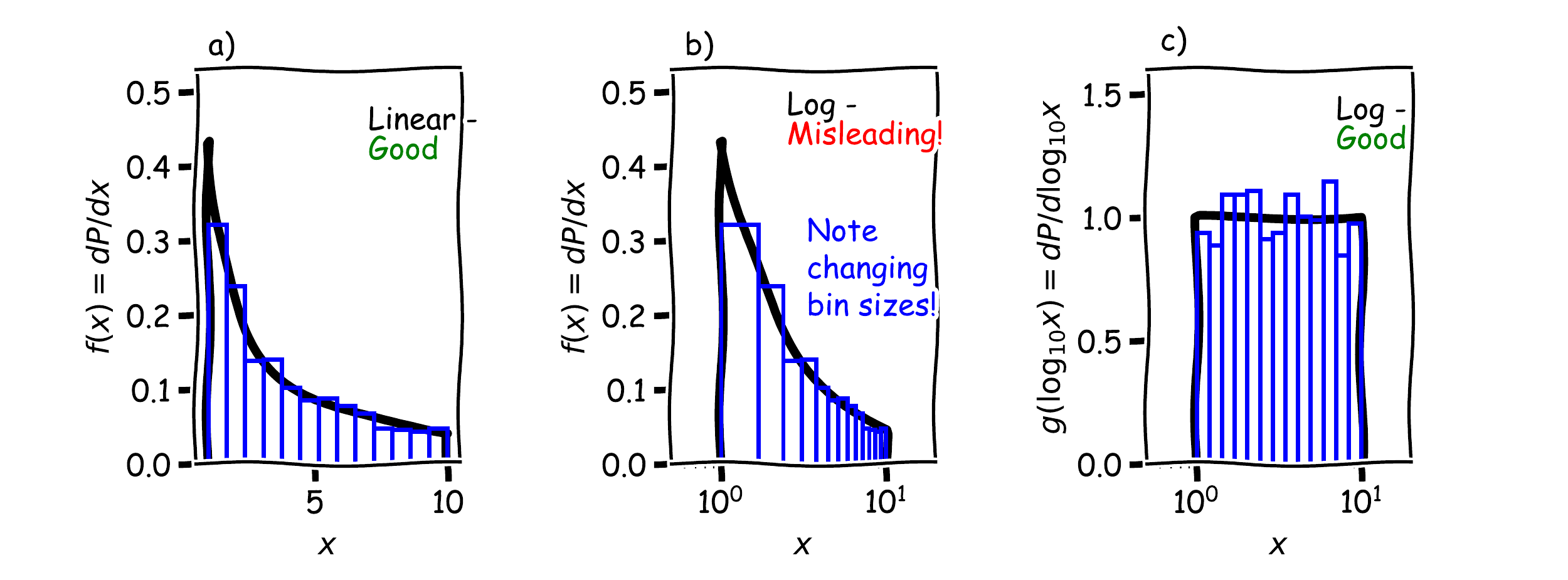}
\caption{ The pedagogical example from section 2. The three panels show three different ways one could imagine plotting a log-uniform probability density. Also shown are normalized histograms analogous to each curve. Panel c shows the distribution in log-space, i.e. where this distribution is uniform. Panel a shows the same distribution in linear space. Panel b shows what happens if you take the code that produces panel a, and simply set the x-axis to be logarithmic without changing anything else. This produces a plot that visually suggests one is more likely to, for instance, draw a value within a factor of 2 of $x=1$, than within a factor of 2 of $x=10$. This is wrong - these two probabilities are equal, as suggested in panel c. From the overplotted histogram, we can see that this can be understood as implicitly using changing bin sizes.}
\label{fig:example}
\end{figure*}

Consider a random variable $x$ whose PDF $f(x) \propto 1/x$ between $a$ and $b$ with $0<a<b$. The normalized PDF is therefore
\begin{equation}
f(x) = \begin{cases}
0\ &\mathrm{for}\ x<a\ \mathrm{or}\ x>b \\
\frac{1/x}{\ln(b/a)}\ &\mathrm{for}\ a\le x \le b
\end{cases}
\end{equation}
From this, we can compute the distribution of the random variable $u=\log_{10} x$, which will become useful momentarily. Defining\footnote{Instead of defining a new function, e.g. $f$ or $g$, for every random variable, one could simply denote them all with e.g. $p$, so we would have $p(x)$ and $p(u)$. Slightly more formally these would be denoted $p_X(x)$ and $p_U(u)$, but the subscript is often suppressed. For clarity, I'll just use $f$ and $g$ for now.} $g(u)$ to be the PDF of $u$, under a change of variables $g(u) = f(x)|\partial x/\partial u|$, so 
\begin{equation}
g(\log_{10} x) = g(u) =  \begin{cases}
0\ &\mathrm{for}\ x<a\ \mathrm{or}\ x>b \\
\frac{\ln 10}{\ln(b/a)} = \mathrm{const.}\ &\mathrm{for}\ a\le x \le b
\end{cases}
\end{equation}

\begin{figure*}
\centering
\includegraphics[width=7.5in]{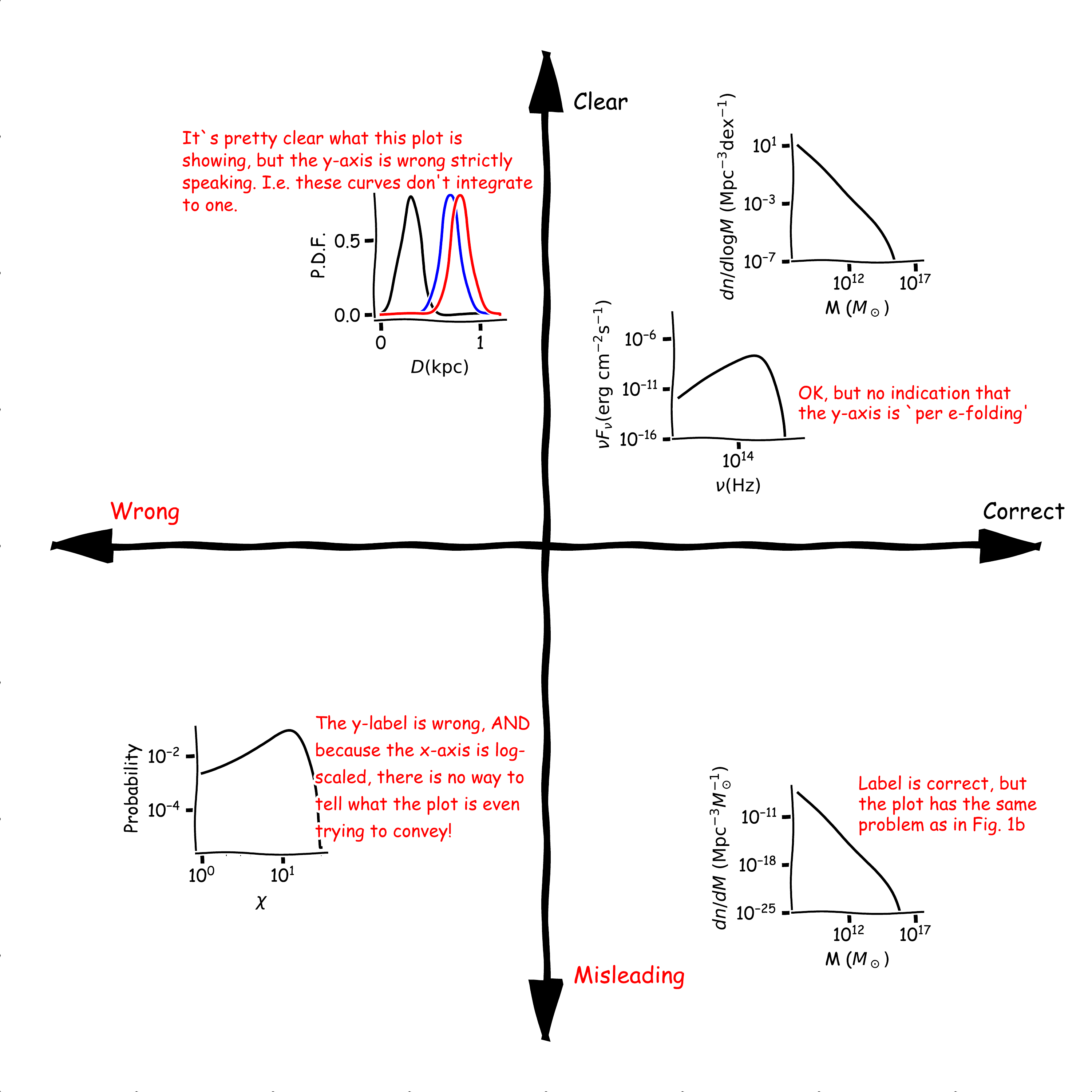}
\caption{ A set of annotated plots showing some of the pitfalls discussed in this work. Plots further to the right are more correct, and plots further up are clearer in their meaning. These axes don't necessarily align, since a plot can be labelled correctly but still be misleading, and a plot can be labelled incorrectly even if the author's point is quite clear. }
\label{fig:roguesgallery}
\end{figure*}

Figure 1 shows a number of ways that this straightforward distribution might be plotted, specifically for $a=1$ and $b=10$. All of the labels on the $y-$axes are correct, avoiding some of the pitfalls mentioned in the introduction. Panels a and c correspond well with our intuition given their respective $x-$axes -- in linear space, values of $x$ close to 1 are more common (per unit $x)$, while in logarithmic space the distribution is uniform - values of $\log_{10}$ between 0 and 1 are equally likely (per unit $\log_{10} x$). The middle panel might lead us to expect that values of $\log_{10}$ close to 0 are more common (per unit $\log_{10}$x) than those close to 1, which is not the case. The source of this confusion is of course that for densities our intuition expects the quantity on the $y-$axis to be ``per unit $x-$axis,'' whereas in panel b the quantity on the $y-$axis is per unit $x$, not the actual quantity spaced evenly on the $x-$axis, namely $\log_{10}x$

To understand visually why panel b is misleading, let's consider a sample of 1000 points drawn from the distribution $f(x)$. If we bin the samples such that the histograms (shown in blue in Figure 1) approximate the various plots of $f$ and $g$, we see that panels a and c have bins that are uniformly-spaced, whereas panel b has bins that are larger at lower values of $x$. No one would plot a histogram with non-uniform bins, but that is precisely what is done implicitly in plots of $dP/dx$ vs $\log_{10} x$.

Figure 2 shows several more examples of plots of densities one may encounter in astronomy. The plots are shown according to how clear they are, and how correct they are. These are slightly different things -- plots may be technically incorrect even if it is clear what the author intended, and plots may technically have the correct labels, but be misleading for the reasons we discuss here (as in Fig. 1b).

\section{Anticipated FAQ}
{\bf In my subfield, everyone plots X vs Y in a way I guess you'd object to. Should I really fight with my coauthors over this?}\\ \\
Probably not! There are certainly situations where avoiding the problem illustrated in Figure 1 might cause more confusion than it would avoid. \\ \\

\noindent
{\bf Hang on, don't you plot $\Sigma$ vs. r all the time? What gives?}\\ \\
Wow, I'm flattered you're familiar with my work! I do often plot the surface density of mass, $\Sigma$, in disks as a function of cylindrical radius $r$, whereas the differential amount of mass per unit radius is actually $2\pi r \Sigma$. This partly falls into the case covered in the previous question, i.e. everyone who studies the density distribution of disks plots $\Sigma$ vs. $r$. In addition though, $\Sigma$ is a physically meaningful quantity, related directly, for example, to the self-gravity of the disk or the expected star formation rate. On top of that, $\Sigma$ itself often has an exceptionally simple form for as-yet poorly-understood reasons, namely $\Sigma \propto \exp(-r/r_s)$, where $r_s$ is some scale-length. This leads us to conclude that even though some things may be interpreted as densities, the differential distribution of mass (literal or probability) is not always the most important point to convey in a plot. \\ \\

\noindent
{\bf Fine, I'll give you $\Sigma$ vs $r$, but shortly after you wrote this ``paper,'' weren't you a co-author on some work plotting power spectral density (PSD) per unit frequency vs. log-frequency, exactly analogous to the problematic panel in Figure 1?} \\ \\

Yes, and I did raise this exact issue with my co-authors. We decided that this also falls under the first case raised in this FAQ. In particular, one of people's few intuitions for power spectra is that a white noise power spectrum should be flat, i.e. have a constant PSD. If we were to make the adjustment I advocate for in this work, a plot of a white noise power spectrum would not be flat, but rather would be inversely proportional to the frequency. To avoid conflicting with people's intuition, i.e. to keep white noise spectra flat, we kept the plots of power per frequency vs. log frequency. So I tried, but not that hard! \\ \\

\noindent
{\bf I'm looking at Figure 2, and I don't understand what you have against $\nu F_\nu$ plots. I thought they were intended to address exactly the problem you're pointing out.} \\ \\

Yes, $\nu F_\nu = \nu dF/d\nu = dF/d\ln\nu$, so one can easily see by eye in a $\nu F_\nu$ plot against $\log \nu$ where most of the flux in the spectrum is being emitted. My only objection (which is why it's closer to the origin in Figure 2, but still in the upper-right quadrant) is that $dF/d\ln\nu$ is not quite the same as $dF/d\log\nu$. The difference is just a constant factor of $\ln 10$, and one usually doesn't care too much about the normalization of these plots. Nonetheless I would personally prefer if the units were ergs per second per square centimeter {\bf per dex}, as opposed to ergs per second per square centimer per e-folding of $\nu$ (the de facto units of $\nu F_\nu$), or ergs per second per square centimer (the not-quite-right label people often use on plots of $\nu F_\nu$). \\ \\

\noindent
{\bf So what's your opinion of plots of Janskys vs. $\log \lambda$?} \\ \\

Let's just say they would be off the chart in Figure 2. \\ \\

\noindent
{\bf Is this what you've been working on instead of responding to my email?} \\ \\

I actually wrote most of this in 2017, so if you've been waiting on a reply for that long, sorry! You should probably ping me again. \\ \\

\noindent
{\bf How did you make the plots look cartoonish?} \\ \\
\noindent
import matplotlib.pyplot as plt\\
with plt.xkcd(): \\
\# usual plotting code here.

\section{Takeaway points}

This is a quick summary of the points I tried to raise. \\ \\

\noindent
-- Probability and Probability Density are different quantities with different units.\\ \\
-- When you're plotting a PDF, remember that the units on the y-axis do have dimensions in general. In particular, they should be something like ``probability per unit-whatever-is-on-the-x-axis.'' If the thing on the x-axis is logarithmic, your y-axis should probably be something per dex. \\ \\
-- There are many exceptions to the latter point. Clarity and not confusing your readers is more important than whether I am personally annoyed by your plot.\\ \\
-- When plotting something that is very similar to a histogram against a continuous variable, it's rare that the y-axis should be ``Probability.'' \\ \\
-- If you're plotting a PDF, you can usually be more explicit with your label than just ``PDF.'' In particular, make sure there is no ambiguity about exactly which variable's PDF is being plotted.  \\ \\
-- Make sure that your PDFs integrate to one. \\ \\

\section*{Acknowledgements}
To be submitted to the Astro-Pedantic Journal on April 1, 2020 for a bit of levity in these unsettling times.

\bibliography{/Users/jforbes/updatingzotlib}

\clearpage

\end{document}